\documentclass[letter,structabstract]{aa}
\usepackage{graphicx}
\usepackage{txfonts}
\usepackage{natbib}

\begin{document}
%
%    \title{CoRoT\thanks{The CoRoT space mission was developed and is operated by the French space agency CNES, with participation of ESA's RSSD and Science Programmes, Austria, Belgium, Brazil, Germany, and Spain.}  opens a new gate to probe red-giant populations in the galactic disk}

    \title{Evidence for a sharp structure variation inside a red-giant star}

    %\title{The CoRoT\thanks{The CoRoT space mission was developed and is operated by the French space agency CNES, with participation of ESA's RSSD and Science Programmes, Austria, Belgium, Brazil, Germany, and Spain.}  %reveals a new gate to access the properties of populations of red giants in the Galaxy}
    %red clump}
  \author{A. Miglio\inst{1}\fnmsep\thanks{Postdoctoral Researcher, Fonds de la Recherche Scientifique - FNRS, Belgium}
          \and
          J. Montalb\'an\inst{1}
          \and
          F. Carrier \inst{2}
          \and
          J. De Ridder\inst{2}
          \and
          B. Mosser\inst{3}
          \and
          P. Eggenberger\inst{4}
          \and
          R. Scuflaire\inst{1}
          \and
          P. Ventura\inst{5}
          \and
          F. D'Antona\inst{5}
          \and
          A. Noels\inst{1}
          \and
          A. Baglin\inst{3}
          }
   \offprints{A.~Miglio, \email{a.miglio@ulg.ac.be}}

   \institute{Institut d'Astrophysique et de G\'eophysique de l'Universit\'e de Li\`ege,
All\'ee du 6 Ao\^ut, 17 B-4000 Li\`ege, Belgium
              \and
             Instituut voor Sterrenkunde, K.U. Leuven, Celestijnenlaan 200D, 3001 Leuven, Belgium
         \and
         LESIA, UMR8109, Universit\'e Pierre et Marie Curie, Universit\'e Denis Diderot, Observatoire de Paris, 92195 Meudon,
France
         \and
                      Observatoire de Gen\`eve, Universit\'e de Gen\`eve, 51 chemin des Maillettes, 1290 Sauverny, Switzerland
         \and
         INAF- Osservatorio di Roma, via Frascati, 33, Monteporzio Catone, RM, Italy
         }

   \date{Received; accepted}

% \abstract{}{}{}{}{}
% 5 {} token are mandatory

  \abstract
   {The availability of precisely determined frequencies of radial and non-radial oscillation modes in red giants is finally paving the way for detailed studies of the internal structure of these stars.}
   {We look for the seismic signature of regions of sharp structure variation in the internal structure of the CoRoT target HR7349.}
    {We analyse the frequency dependence of the large frequency separation and second frequency differences, as well as the behaviour of the large frequency separation obtained with the envelope auto-correlation function.}
    %results
    {We find evidence for a periodic component in the oscillation frequencies, i.e. the seismic signature of a sharp structure variation in HR7349. In a comparison with stellar models we interpret this feature as caused by a local depression of the sound speed that occurs in the helium second-ionization region. Using solely seismic constraints this allows us to estimate the mass ($M=1.2^{+0.6}_{-0.4}$ $M_\odot$) and radius ($R=12.2^{+2.1}_{-1.8}$ $R_\odot$) of HR7349, which agrees with the location of the star in an HR diagram. }
    %, i.e. low-mass stars that passed the He-flash and are in the core He-burning phase
  %  {We have shown that asteroseismology of red-giant stars opens the possibility of deriving detailed information on the internal structure of these evolved stars.}
  {}
   \keywords{stars: fundamental parameters -- stars: individual: HR7349 -- stars: interiors -- stars: late-type -- stars: oscillations}

   \maketitle
%
%________________________________________________________________

\section{Introduction}
\label{sec:intro}
Red giant stars are cool, highly luminous stars. They play a predominant role in stellar, galactic, and extragalactic astrophysics because they serve as distance and age indicators for globular clusters and nearby galaxies \citep[see e.g.][]{Lee93, Girardi01}. Moreover their observed chemical composition is a fundamental ingredient in the study of the chemical evolution of their host galaxy \citep[see e.g.][ and references therein]{Catelan09}.

%These stars are old, they have undergone their phase of core hydrogen burning, followed by a contraction of their helium core and a drastic expansion of their outer parts. Their precise evolutionary state is however very difficult to ascertain: while showing similar surface global properties (namely brightness and color), they can be either burning hydrogen in a shell located above the helium core, or in a readjustment phase following the onset of core helium burning, or in a quiet phase of core helium burning, or in a post-He-burning phase \citep[see e.g.][for a recent review]{Salaris02}. Their mass is even more difficult to derive from these global properties since stars in a wide domain of mass (0.8 to 3 solar masses) fulfill similar global properties at one or another of these evolutionary stages.

% red giants obs.
All red giants have an extended and diluted convective envelope surrounding a dense core, which makes their structure and therefore their pulsation properties very different from that of our Sun. As in solar-like stars however, their convective envelope can stochastically excite acoustic oscillation modes.
While stochastic oscillations have been firmly detected in a few bright G and K giants since \citet{Frandsen02}, a major breakthrough was achieved thanks to the photometric space-based observations with the CoRoT\footnote{The CoRoT space mission, launched on December 27th 2006, has been developed and is operated by CNES, with the contribution of Austria, Belgium, Brazil, ESA (RSSD and Science Programme), Germany and Spain.} satellite \citep{Baglin06}, leading to the unambiguous detection of radial and non-radial oscillation modes in thousands of red giants \citep{DeRidder09, Hekker09, Carrier10, Mosser10}.
Moreover, the \textit{Kepler} space-telescope has also started monitoring red giants for photometric variations and has already detected radial and non-radial oscillation modes in a large number of field giants as well as cluster members \citep{Bedding10, Stello10}.  As the precise detection of oscillation modes in the Sun fostered helioseismic inferences in the 1990s, we now have the chance to start probing the internal structure of these cool evolved stars thanks to the recent advances in observational stellar seismology.

% glitches
Stellar global acoustic oscillation modes show up in the frequency spectrum as discrete peaks.  While the average spacing between the frequencies $\nu$ of modes of the same spherical degree (large frequency separation, $\Delta\nu$) represents an estimate of the mean density of the star \citep{Vandakurov67, Tassoul80}, periodic deviations from a constant $\Delta\nu$ are signatures of so-called acoustic glitches, i.e. regions of sharp-structure variation in the stellar interior.
Any localised region of sharp variation of the sound speed $c$ induces in the frequencies an oscillatory component with a periodicity related to the sound-travel time measured from that region to the surface of the star \citep{Vorontsov88,Gough90}. Its amplitude depends on the sharpness of the glitch and decays with $\nu$, because as the frequency increases, the wavelength of the mode becomes comparable with or less than the extent of the glitch.  These sharp variations are found in zones of rapidly changing chemical composition, in ionization zones of major chemical elements, or in regions where the energy transport switches from radiative to convective.

In the Sun, the detection and interpretation of a large number of acoustic oscillation modes led to the determination of the depth of both the solar convective envelope and of the helium second-ionization zone, and allowed us to estimate the helium abundance in the envelope, otherwise inaccessible by other means \citep[see][for a review]{jcd02}.

The possibility and relevance of detecting sharp structure variations in other main-sequence stars, as well as first attempts to look for these signatures, have been addressed in several papers in the literature \citep[see e.g. ][]{Perez98, Roxburgh98b,Monteiro98,Monteiro00,Mazumdar01,Miglio03,Ballot04,Basu04,Theado05,Mazumdar05,Verner06,Houdek07a,Bedding10a}.

Here we report on the detection of a sharp-structure variation in the red giant CoRoT target HR7349 (HD181907), a star with a physical structure very different from the Sun, yet showing global oscillations modes that allow us to probe its internal structure.

%\begin{figure}
%\centering
%   \includegraphics[width=\hsize]{figure/012ls}
%      \caption{Large separation computed with $\ell=0, 1$ and $2$ frequencies listed by \citet{Carrier10}; a linear trend was removed from $\Delta\nu$. The frequency of the fitted component (dashed line) is found to be $0.13 \pm 0.02$ Ms, which allows us to infer the relative acoustic radius ($t_0/T = 0.55 \pm 0.07$) of the acoustic glitch.}
%         \label{fig:fig1}
%\end{figure}
%

%----------------------------------------------------------------
\section{\bf{Evidence for} an oscillatory component in the frequency spacings}
%----------------------------------------------------------------
\citet{Carrier10} have determined the eigenfrequencies of HR 7349.
They have shown that the first frequency differences
$\Delta\nu_{n,\ell}\equiv\nu_{n,\ell}-\nu_{n-1,\ell}$ are
modulated. In order to investigate this modulation, we have
analysed these first frequency differences and compared them to
the function $\Delta\nu(\nu)$ derived from the method developed by
\citet{Mosser09}. This envelope autocorrelation function (EACF) is
obtained from the auto-correlation of the times series, performed
as the Fourier spectrum of the windowed Fourier spectrum; the
windows are Hanning filters covering the frequency range where
significant excess power is observed. Because the modulation is
rapid, we had to use narrow filters, with a FWHM varying between 1
and 2 times the mean value of the large separation. An EACF
computed with a narrow filter benefits from the low filtering, but
suffers from larger error bars.

\begin{figure}
\centering
   \includegraphics[width=.85\hsize]{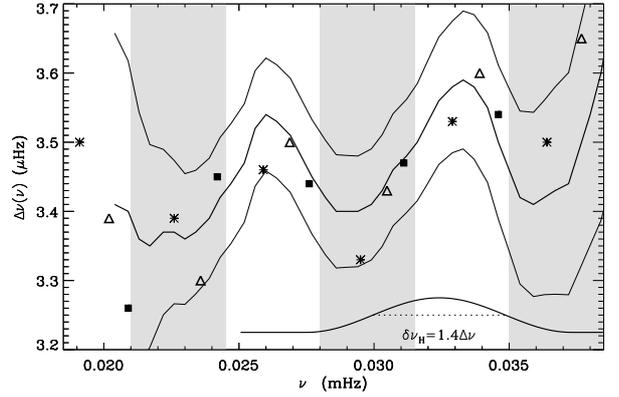}
      \caption{Large separations $\Delta\nu_{n,\ell}$ computed by \citet{Carrier10}  ($\ell=0$: black squares; 1: stars; 2: triangles) are superimposed on the $\Delta\nu_{n,\ell}$ function given by the autocorrelation of the time series (solid thick line). The $\pm$1-$\sigma$ error bars are given by the two adjacent curves; they agree with the error bars of $\Delta\nu_{n,\ell}$. Both $\Delta\nu_{n,\ell}$ and $\Delta\nu_{n,\ell}$  show a periodic modulation with a period of the order of 0.14 Ms. The zebra regions indicate 1-$\Delta\nu$ wide frequency ranges. The inset shows the filter used for performing the envelope autocorrelation function, with a full-width at half-maximum $\delta\nu_{\mathrm{H}}$ equal to 1.4 times the mean value of the large separation.}
         \label{fig:fig1}
\end{figure}

Fig.~\ref{fig:fig1} shows the large separations for $\ell=0, 1$,
and $2$ superimposed on the function $\Delta\nu(\nu)$ obtained
with a filter width equal to 1.4 times the mean value of the large
separation, which represents an acceptable compromise between the
filtering of the signal and the noise level. We remark that both values
and error bars of $\Delta\nu(\nu)$ and $\Delta\nu_{n,\ell}$ closely agree.
The
modulation seen in the function $\Delta\nu(\nu)$ is present in all
autocorrelations performed with a filter that is narrow enough to respect
the Shannon criterion.

We have detrended the low-frequency gradient of $\Delta\nu(\nu)$
in order to emphasize the modulation, which is of the order
of $7\,\mu$Hz (a period of about $\mathcal{T}_0 = 0.14$ Ms).

That the modulation is seen with two independent methods,
is present over more than 2.5 periods, and remains detectable even
if a significant level of noise is added to the spectrum makes us
confident that the
oscillation found in the frequency differences is genuine.

The uncertainty of the period $\mathcal{T}_0$ of the oscillatory component was estimated with a Monte Carlo simulation. We assumed that the frequencies are independent and are Gaussian-distributed around their observed value with a standard deviation equal to their 1-$\sigma$ error bar as listed by \citet{Carrier10}. We then generated 5000 realizations of the frequencies, computed the frequency differences, and applied a non-linear regression procedure.
Each time a cost function $S$, defined as
\begin{equation}
S(\beta)=\left(y-\mu(\beta)\right)^t V^{-1}\left(y-\mu(\beta)\right)
\end{equation}
was minimized, where $y$ is the matrix of observed frequency differences, $V$ is the covariance matrix of the observed frequency differences, and $\mu(\beta)$ is the matrix containing the non-linear model
\begin{equation}
\mu(\beta; \nu) = a + b \nu + A \sin(2\pi\, \mathcal{T}_0 \nu) + B \cos(2\pi\, \mathcal{T}_0 \nu)
\end{equation}
with $\beta \equiv (a,b,A,B,\mathcal{T}_0)$ the fit parameters, and $\nu$ the covariates. The uncertainty of $\mathcal{T}_0$ was then derived as the standard deviation of the sample of $\mathcal{T}_0$ values. As a result, we obtain a period $\mathcal{T}_0=0.13 \pm 0.02$ Ms from the component detected in $\Delta\nu$  and $\mathcal{T}_0=0.14 \pm 0.03$ Ms when fitting the behaviour of the second frequency differences $\Delta_2 \nu$ \citep[as defined in][]{Gough90}.

\section{Interpretation of the oscillatory component}
%----------------------------------------------------

As recalled in Sect. \ref{sec:intro}, the periodicity of the component allows us to infer the location of the glitch in terms of its acoustic depth or in terms of its relative acoustic radius (as suggested by \citealt{Ballot04}). For HR7349 we find: $t_0/T = 0.55 \pm 0.07$, where $t$ is the sound travel time from the centre to the glitch, and $T$ is the acoustic radius of the star ($T=(2\left<\Delta\nu\right>)^{-1}$).

\begin{figure}
\centering
   \includegraphics[width=.9\hsize]{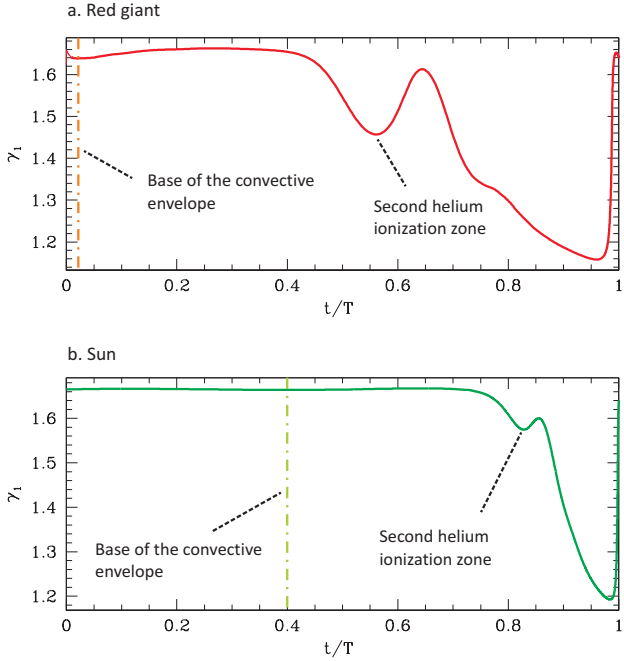}
      \caption{$a)$ Behaviour of $\gamma_1$ as a function of $t/T$ in a 1.2 $M_\odot$, 12 $R_\odot$ model: the region where helium undergoes its second ionization induces a local minimum in $\gamma_1$. In the same model the base of the convective envelope is located at $t/T = 2\; 10^{-2}$. $b)$ Location of acoustic glitches in a model of the Sun.}
         \label{fig:gamma1}
   \end{figure}

To interpret this observational evidence, we consider in Fig. \ref{fig:gamma1} the location of acoustic glitches in the structure of red-giant models computed with the code ATON \citep{Ventura08}.
We find that at about half the acoustic radius the sound speed has a sharp variation owing to a local depression of the first adiabatic exponent\footnote{$\gamma_1=\left(\partial \ln{p}/\partial \ln{\rho}\right)_s$, where $p$,  $\rho$, and $s$ denote pressure, density, and specific entropy.}  $\gamma_1$ in the helium second-ionization zone. We therefore interpret the periodic component detected in the frequencies of HR7349 as the signature of helium second ionization.
This interpretation is further supported by a direct comparison between observed and numerically computed pulsation frequencies in models of red giants (see Sect. \ref{sec:models}).

We notice that while in the Sun and solar-like stars the base of the convective envelope is located at about half the stellar acoustic radius (see Fig. \ref{fig:gamma1}), in red giants it is located at a small relative acoustic radius (typically less than $10^{-1}$ for the models considered, depending on the mass and evolutionary state).
%In this respect, red giants are favourable cases to detect the He-ionization signature, provided that sufficient relative precision in the frequencies is achieved.

\begin{figure}
\centering
   \includegraphics[width=.9\hsize]{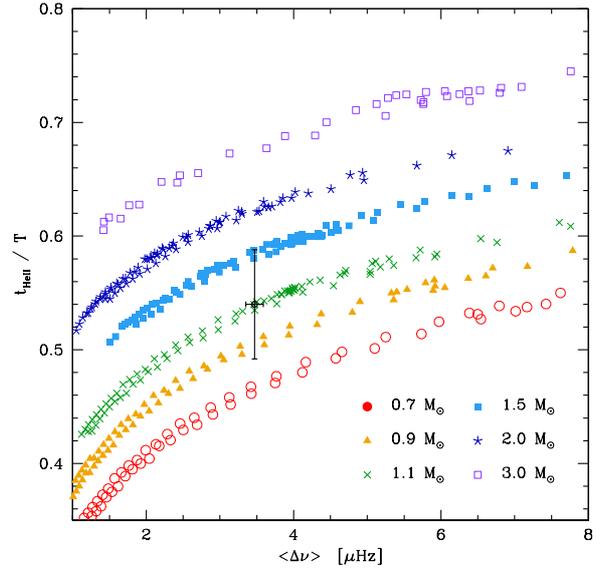}
      \caption{Behaviour of the relative acoustic radius of the second helium ionization region ($t_{\rm HeII}/T$) as a function of the average large frequency separation in stellar models with different mass (0.7 to 3 $M_\odot$), initial chemical composition (helium mass fraction $Y=0.25, 0.278$, heavier-elements mass fraction $Z=0.02, 0.01$), and convection-efficiency parameter ($\alpha_{\rm MLT}=1.6, 1.9$). The position of HR7349 in this diagram is represented by the black dot with error bars.}
         \label{fig:fig4}
   \end{figure}

Furthermore, we find that in the models the relative acoustic radius of the He second-ionization region ($t_{\rm HeII}/T$) and the average $\Delta\nu$ determine the mass and radius of the star (see Fig. \ref{fig:fig4}). We ascribe this dependence to the the fact that the temperature stratification in the envelope of red-giant stars depends mostly on $M$ and $R$ \citep[see e.g.][]{Kippenhahn}; a thorough discussion on this topic will be reported in detail in a forthcoming paper.

On the basis of seismic constraints only, we can thus derive from Fig. \ref{fig:fig4} the mass of HR7349 ($M= 1.2^{+0.6}_{-0.4}$ $M_\odot$) and, through the scaling of $\Delta\nu$ with the mean stellar density, its radius ($R= 12.2^{+2.1}_{-1.8}$ $R_\odot$).

Because the parallax of this nearby red giant is known, an  independent radius estimate of HR7349 can be obtained via the precise absolute luminosity and effective temperature \citep{Carrier10}, leading to $R = 12.2 \pm 1.1$ $R_\odot$, in agreement with the seismically inferred radius. We also notice that the mass and radius estimates are also compatible to those obtained by the combination of the frequency corresponding to the maximum oscillation power ($\nu_{\rm max}$), $\left<\Delta\nu\right>$, and $T_{\rm eff}$ \citep[][]{Carrier10}.

\section{Direct comparison with numerically computed oscillation frequencies}
\label{sec:models}
We consider a red giant model of 1.2 $M_\odot$ and 12 $R_\odot$ computed with the stellar evolutionary code ATON \citep{Ventura08}. The model has an initial helium mass fraction $Y=0.278$ and a heavier-elements mass fraction $Z=0.02$. For this model we computed adiabatic oscillation frequencies of spherical degree $\ell =0,1$ and 2 in the domain 15-40 $\mu$Hz with an Eulerian version of the code LOSC \citep{Scuflaire2008b}.

\begin{figure}
\centering
   \includegraphics[width=.9\hsize]{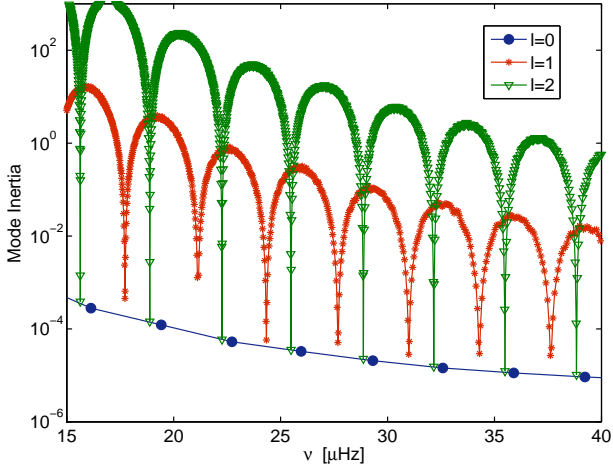}
      \caption{Mode inertias of radial modes (full dots), $\ell = 1$ modes (asterisks) and $\ell=2$ modes (open triangles) for a 1.2 $M_\odot$ red-giant model.}
         \label{fig:fig5}
   \end{figure}

In a red giant model, differently from the Sun, non-radial oscillation modes in the frequency domain of solar-like oscillations \citep{Kjeldsen95} are expected to have a mixed pressure and gravity nature. In the model we considered (see Fig. \ref{fig:fig4})  the spectrum of $\ell=1$ and 2 modes is mostly populated by modes with a dominant gravity-like behaviour. These modes exhibit a high inertia compared to radial modes, and therefore show surface amplitudes significantly smaller than the purely acoustic radial modes \citep{Dziembowski01, jcd04, Dupret09, Eggenberger10}.
Nonetheless, among this large number of non-radial oscillation modes, few modes are efficiently trapped in the acoustic cavity of the star and have inertias similar to those of radial modes (see Fig. \ref{fig:fig5}).

\begin{figure}
\centering
   \includegraphics[width=.9\hsize]{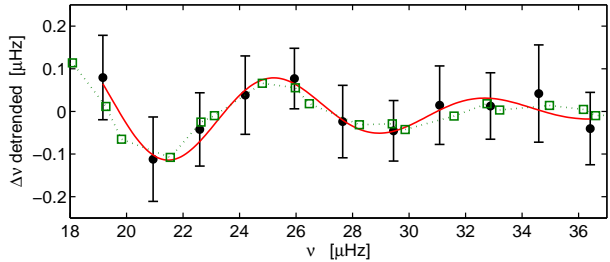}
      \caption{Open squares represent $\Delta\nu_{n,\ell}$ computed from $\ell=0,1,2$ adiabatic frequencies in a 1.2 $M_\odot$ red giant model. The solid line shows a sinusoidal component with amplitude decreasing with frequency \citep[see][]{Monteiro98,Houdek07a} fitted to the $\ell=0,1$ large separation determined by \citealt{Carrier10} (dots with error bars).}
         \label{fig:model}
   \end{figure}

We then compute $\Delta\nu(\nu)$ and $\Delta_2\nu(\nu)$ by selecting radial modes and the $\ell=1$ modes corresponding to local minima in the inertia, hence strongly trapped in the acoustic cavity. We find an average value of the large frequency separation of 3.4 $\mu$Hz and clear periodic components in $\Delta\nu(\nu)$ (open squares in Fig. \ref{fig:model}) and $\Delta_2\nu(\nu)$.

The comparison between the model and observed frequency separations clearly supports the interpretation that the periodic component is caused by the local depression in the sound speed in the second helium ionization zone. This feature is well represented by current models that have a mass and radius determined from the periodicity of the component and the average value of the large frequency separation (see Fig. \ref{fig:fig4}).

We notice that the periodic component related to the He second-ionization zone is expected to have an amplitude decaying with frequency, as shown in numerically computed frequencies (open squares in Fig. \ref{fig:model}) and in analytical approximations \citep{Monteiro98,Houdek07a}. Although current error bars are too large to draw a firm statement, the amplitude of the observed component computed with $\ell=0$ and 1 modes seems to show the expected behaviour, while we notice that the observed $\ell=2$ modes show a larger scatter.

\section{Conclusions}
In summary, we have shown that the acoustic pulsation modes detected by CoRoT in the red giant star HR7349 bear the signature of a sharp-structure variation inside the star. Comparison with stellar models allows us to interpret this feature as caused by a local depression of the sound speed occurring in the helium second-ionization region.

Besides representing the first seismic inference of a local feature in the internal structure of an evolved low-mass star, this detection allows a mass ($M=1.2^{+0.6}_{-0.4}$ $M_\odot$) and radius ($R=12.2^{+2.1}_{-1.8}$ $R_\odot$) estimate based solely on seismic constraints. Moreover, for this nearby star, we could also check that our radius estimate is compatible with the one based on luminosity and effective temperature ($R = 12.2 \pm 1.1$ $R_\odot$). This additional test reinforces the proposed interpretation and approach, which could be applied to the thousands of pulsating giants of unknown distance that are currently being observed with the space satellites CoRoT (see \citealt{Hekker09} and, in particular, Fig. 6 in \citealt{Mosser10}) and \textit{Kepler} \citep{Boruki10, Gilliland10, Bedding10}. A reliable seismic estimate of the mass and radius of these stars would represent an essential ingredient for stellar population studies \citep{Miglio09} and for characterizing planets orbiting around these evolved distant stars \citep[see e.g.][]{Hatzes07}.

We finally recall that the detectability of the signature of He ionization can potentially lead us to a seismic estimate of the envelope helium abundance in old stars.
Indeed, as shown for the Sun and solar-like stars, the amplitude of the periodic component depends on the envelope helium abundance \citep[see e.g.][and references therein]{Basu04, Houdek07a}. While CoRoT and \textit{Kepler} observations will provide other targets and further reduce the uncertainty on the oscillation frequencies, a thorough study on the required precision in terms of seismic and non-seismic observational constraints, as well as and in terms of models, should be undertaken to aim for reliable seismic estimate of the envelope helium abundance in giants.

%Basu: This opens up the possibility of using asteroseismic data to obtain the helium abundance in stellar envelopes, leading to a better understanding of the evolution of stellar populations in our galaxy.

%The helium abundance of stars is often used to extrapolate back to obtain estimates of the primordial helium abundance, which is an important test of cosmological models. The helium content of the oldest stars cannot however, be measured directly Ð these are low mass stars, and their photospheres are not hot enough to excite helium lines that can be used to determine the helium abundance spectroscopically.

\begin{acknowledgements}
JM acknowledges the Belgian Prodex-ESA for support (contract C90310). FC is a postdoctoral fellow of the Funds for Scientific Research, Flanders (FWO).
\end{acknowledgements}

\bibliographystyle{aa}
\small
\bibliography{../andrea}
%
%\begin{thebibliography}{}
%
%\end{thebibliography}

\end{document}